\documentclass[twocolumn,amssymb, aps, prb,showpacs,superscriptaddress]{revtex4}
\usepackage{bm}
\usepackage{graphicx}
\usepackage{color}
\usepackage{amsmath, amssymb}
\usepackage{braket}
\usepackage{color}
\usepackage{comment}
\begin{document}
\title{Switching magnetic texture via in-plane magnetic field in noncentrosymmetric dipolar magnets: From skyrmions to antiskyrmions and nontopological magnetic bubbles}
\author{Tatsuki Muto}
\author{Masahito Mochizuki}
\affiliation{Department of Applied Physics, Waseda University, Okubo, Shinjuku-ku, Tokyo 169-8555, Japan}
\date{\today}
\begin{abstract}
We theoretically investigate field-induced switching of magnetic topology in a nanodisk-shaped sample of noncentrosymmetric dipolar magnet in which the Dzyaloshinskii-Moriya interaction that stabilizes an antiskyrmion with $N_{\rm sk}$=+1 and the magnetic dipole interaction that stabilizes a skyrmion with $N_{\rm sk}$=$-1$ are in keen competition where $N_{\rm sk}$ is the skyrmion number. Our micromagnetic simulations demonstrate that the competition offers a unique opportunity to switch magnetic textures with distinct magnetic topology among the antiskyrmion ($N_{\rm sk}$=+1), elliptical skyrmion ($N_{\rm sk}$=$-1$), and nontopological bubble ($N_{\rm sk}$=0) in a deterministic manner by application of magnetic fields parallel to the sample plane. By calculating time and spatial profiles of energy contributions from respective interactions and magnetic anisotropy, we clarify the physical mechanism and properties of the observed field-induced topology switching phenomena. Our findings are expected to provide useful insights into the spintronic application of topological magnetism.
\end{abstract}
\maketitle

\section{Introduction}
The experimental discovery of magnetic skyrmions in a small-angle neutron scattering experiment was reported in 2009~\cite{Muhlbauer2009}. Subsequently, their real-space images were observed by Lorentz transmission electron microscopy in 2010~\cite{YuXZ2010}. In the latter experiment, hexagonal crystallization of the vortex-like magnetic skyrmions was observed in thin-plate samples of chiral magnets under application of a magnetic field perpendicular to the sample plane. Since these epoch-making experiments, magnetic skyrmions have attracted enormous research interest continuously as promising building blocks for spintronics devices~\cite{SekiBook2016,Nagaosa2013,Fert2013,Fert2017,Everschor2018,Tokura2020,ZangX2020,YangS2024,Ohki2024,Mochizuki2015}. They have been discovered not only in chiral magnets~\cite{Munzer2010,YuXZ2011,Seki2012a,Seki2012b,Tokunaga2015,Karube2017} but also in various other magnetic systems such as polar magnets~\cite{Kezsmarki2015,Ruff2015,Fujima2017,Akazawa2022,Kurumaji2017,Kurumaji2021}, frustrated magnets~\cite{Okubo2012,Leonov2015,Leonov2017,Hayami2016a,Hayami2016b,ZhangX2017}, magnetic dipolar systems~\cite{Correspondent1972,LinYS1973,Malozemoff1979,Giess1980,Garel1982,Suzuki1983,Hubert1998,YuXZ2012,YuXZ2014,Nagao2013}, spin-lattice coupled magnets~\cite{Muto2023},  centrosymmetric itinerant magnets~\cite{Kurumaji2019,Hirschberger2020a,Hirschberger2020b,Hirschberger2019,Hirschberger2021,Khanh2020,Yasui2020,Hayami2021R}, atomic layers~\cite{Romming2013,Heinze2011,Wiesendanger2016}, and magnetic bilayer heterojunctions~\cite{ChenG2015,Moreau-Luchaire2016,Soumyanarayanan2017}. Furthermore, a variety of topological magnetic structures similar to skyrmions have been experimentally observed or theoretically predicted~\cite{SekiBook2016,Gobel2021}. The magnetization configuration of a skyrmion in two dimensions is characterized by a topological invariant $N_{\rm sk}$ called the skyrmion number, which is defined as,
\begin{align}
N_{\rm sk}=\frac{1}{4\pi} \int_{\rm UC}\bm m \cdot \left(
\frac{\partial \bm m}{\partial x} \times \frac{\partial \bm m}{\partial y}
\right) dxdy,
\end{align}
where $\bm m=\bm m(\bm r)$ is the normalized magnetization vector at position $\bm r$. The spatial integration over $\bm r$ is taken within the magnetic unit cell (UC). This quantity corresponds to a sum of solid angles spanned by neighboring three magnetization vectors $\bm m$ divided by 4$\pi$. In a skyrmion, the magnetization vectors at periphery are oriented parallel to the external magnetic field to maximize the energy gain of the Zeeman interaction, whereas those at center are oriented antiparallel to the magnetic field. Between the periphery and the center, the magnetization vectors gradually rotate as they go along the diameter direction [Fig.\ref{Fig01}(a)]. Because the magnetization vectors of a skyrmion are arranged to cover the surface of unit sphere, the sum of their solid angles takes $\pm 4\pi$ and hence $N_{\rm sk}$=$\pm 1$. If the external magnetic field is positively perpendicular (the magnetization at center is negatively perpendicular) to the skyrmion plane, the sum of the solid angles takes $-4\pi$ and hence $N_{\rm sk}$=$-1$.

The vorticity $n_\Omega$, which is another quantity that characterizes the magnetization configuration, is $+1$ for skyrmions~\cite{Nagaosa2013}. On the other hand, there is another type of topological magnetic structure with an antivortex configuration called antiskyrmion, whose vorticity is $-1$~\cite{Nagaosa2013,Koshibae2016}[Fig.\ref{Fig01}(b)]. The magnetization vectors of antiskyrmion are also arranged to cover the surface of unit sphere. However, when the external magnetic field is positively perpendicular, the sum of their solid angles is $+4\pi$ and $N_{\rm sk}$=+1. Namely, their signs are opposite to those of skyrmions. The antiskyrmions have been recently discovered experimentally in an inverse Heusler alloy Mn$_{1.4}$Pt$_{0.9}$Pd$_{0.1}$Sn~\cite{Nayak2017,Saha2019} and schreibersite compounds (Fe,Ni)$_3$P~\cite{Karube2021,Karube2022,Guang2024}.

Such topological magnetic structures with a finite solid-angle sum generate emergent magnetic fields acting on conduction electrons through exchange coupling with their spins~\cite{Nagaosa2013,Nagaosa2012}. Since the contribution from a single skyrmion or antiskyrmion corresponds to exactly one magnetic flux quantum, a magnitude of the generated emergent magnetic field, which is proportional to the magnetic flux density, can be significantly large ranging from tens to hundreds of Teslas. A typical example of the physical phenomena resulting from the emergent magnetic fields is the Hall effect due to the magnetic topology, which is referred to as the topological Hall effect~\cite{ZangJ2011,Schulz2012}.

The sign of the Hall voltage is reversed when the magnetic field is reversed. Accordingly, if we can reverse or turn off the emergent magnetic field through varying the magnetic topology with external stimuli, it enables us to control the topological Hall effect. Such phenomena are interesting not only for fundamental science but also for technological applications, because the Hall effect is exploited for a wide range of modern electronics devices for, e.g., magnetic-field sensing, current detection, speed measurement, directional control/sensing, and proximity sensing~\cite{Ramsden2006}. The switching of magnetic topology with external stimuli has recently been demonstrated experimentally for certain magnetic materials~\cite{Peng2020,Jena2020,Yasin2023,YuXZ2024,Yoshimochi2024}. The work in Ref.~\cite{Peng2020} demonstrated the controlled switching of magnetic textures among skyrmions ($N_{\rm sk}$=$-1$), antiskyrmions ($N_{\rm sk}$=+1) and nontopological magnetic bubbles ($N_{\rm sk}$=0) in a Heusler alloy Mn$_{1.4}$Pt$_{0.9}$Pd$_{0.1}$Sn by turning an in-plane magnetic field on and off.

In this paper, motivated by this experiment, we theoretically study the field-induced transformations of magnetic textures and switching of magnetic topology in a nanodisk-shaped sample of noncentrosymmetric dipolar magnet in which the Dzyaloshinskii-Moriya (DM) interaction~\cite{Dzyaloshinskii1957,Moriya1960a,Moriya1960b,Fert1980} that favors antiskyrmions with $N_{\rm sk}$=+1 and the magnetic dipole interaction that favors skyrmions with $N_{\rm sk}$=$-1$ are in keen competition~\cite{Karube2021,Karube2022,Peng2020,Heigl2021,Yasin2024}. Our micromagnetic simulations demonstrate that the application of an in-plane magnetic field can switch the magnetic texture among three magnetic states mentioned above in a controlled manner. By investigating time and spatial profiles of energy contributions from several interactions and magnetic anisotropies, we reveal the physical mechanism as well as the dynamical properties of the observed field-induced topology switching phenomena. Our findings are expected to contribute to the development of the future spintronic technology based on topological magnetism.

\section{Model and Method}
The equilibrium magnetic textures that emerge in Mn$_{1.4}$Pt$_{0.9}$Pd$_{0.1}$Sn typically exhibit relatively large sizes on the order of 100 nm with gradual spatial modulation of magnetization vectors. Such magnetic textures can be well described by continuum spin models. However, during the process of field-induced topological switching in this system, magnetization configurations with abrupt spatial variations and local singularities emerge. To accurately describe such magnetization configurations, it is necessary to use an atomistic spin model. In fact, the detailed structures of these spatial magnetization patterns and the presence of singularities significantly affect the stability of magnetic textures and the transformation process. Therefore, in order to faithfully capture the magnetization configurations that appear in the transient process, we employ a lattice spin model in this study.

Specifically, we start with a classical Heisenberg model on a square lattice. The Hamiltonian is given by, 
\begin{align}
\label{H2}
\mathcal{H}
&=-J\sum_{\langle i,j \rangle}\bm m_i \cdot \bm m_j 
\nonumber \\
&+D\sum_i 
(\bm m_i \times \bm m_{i+\hat{x}} \cdot \bm e_x
-\bm m_i \times \bm m_{i+\hat{y}} \cdot \bm e_y)
\nonumber\\
&+I_{\rm dip} \sum_{i<j} \left[\frac{\bm m_i \cdot \bm m_j}{r_{ij}^3}
-\frac{3(\bm m_i \cdot \bm r_{ij})(\bm m_j \cdot \bm r_{ij})}{r_{ij}^5} \right] 
\nonumber \\
&-A\sum_i m_{iz}^2 - B_z \sum_i m_{iz} - \bm B_{\rm in} \cdot \sum_i \bm m_i
\end{align}
where $\bm m_i$ is a classical spin vector at the $i$th site whose norm is unity ($|\bm m_i| = 1$). The first and second terms describe the ferromagnetic exchange interaction and the DM interaction between the nearest-neighbor spins. The third term represents the magnetic dipole interaction, which is long-ranged and acts among all the spin pairs in the system. The fourth term depicts perpendicular magnetic anisotropy with $A(>0)$. The fifth and last terms represent the Zeeman interactions with the perpendicular magnetic-field component $\bm B_{z}=(0,0,B_z)$ and the in-plane magnetic-field component $\bm B_{\rm in}=(B_x,B_y,0)$, respectively.

\begin{figure}[tb]
\includegraphics[scale=1.0]{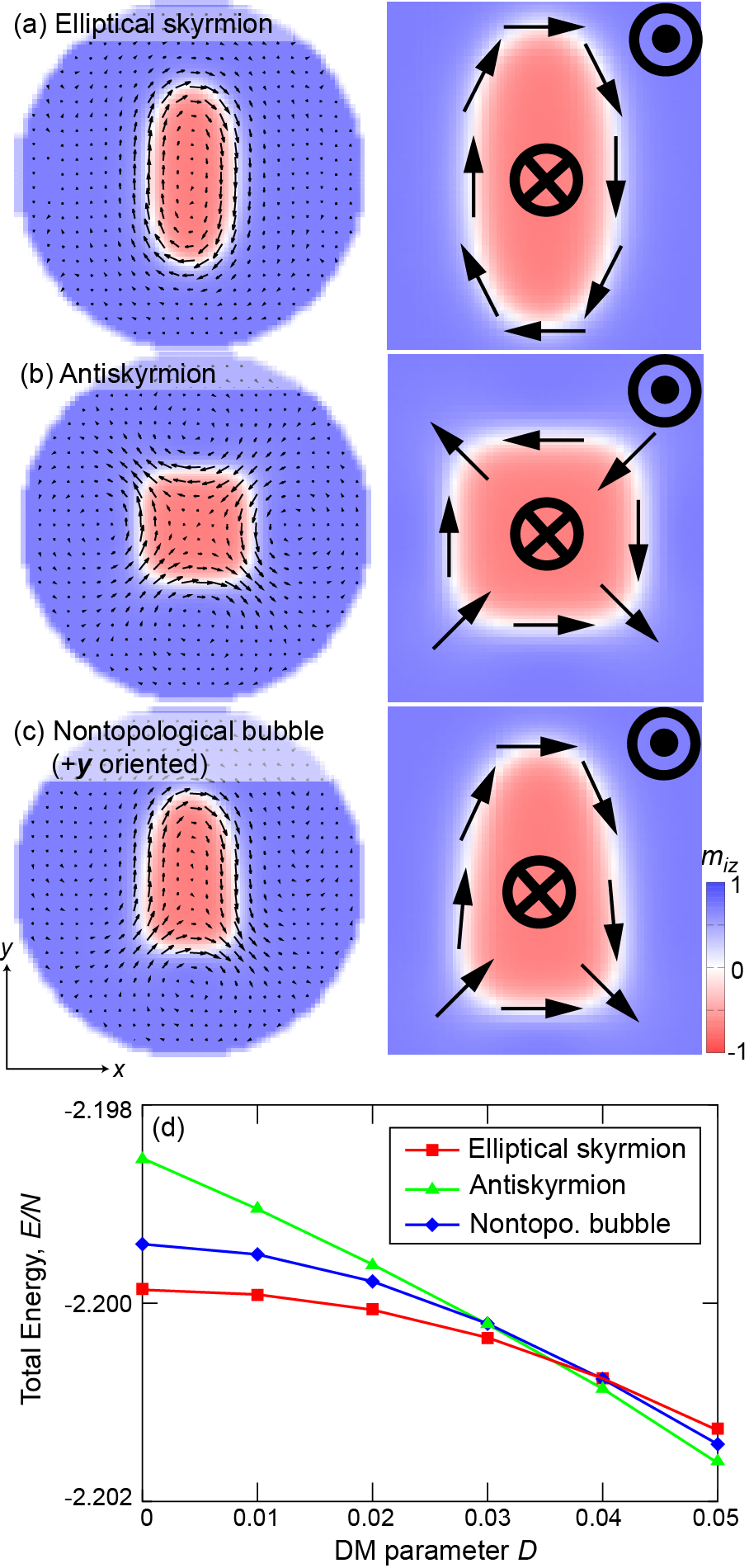}
\caption{(a)-(d) Stable or metastable magnetic structures in a nanodisk-shaped sample of noncentrosymmetric dipolar magnet studied in this work, i.e., (a) elliptical skyrmion, (b) antiskyrmion, and (c) nontopological bubble in a perpendicular magnetic field. Right panels show their schematic illustrations. The arrows depict in-plane components of the magnetization vectors, while the colors represent their out-of-plane components. (d) Calculated energies of these three magnetic structures as functions of the DM parameter $D$.}
\label{Fig01}
\end{figure}
We simulate the field-induced magnetization dynamics by numerically solving the Landau-Lifshitz-Gilbert (LLG) equation using the fourth-order Runge-Kutta method. The equation is given by,
\begin{equation}
\frac{\partial \bm m_i}{\partial \tau}=-\frac{1}{1+\alpha^2} \left[
\bm m_i\times \bm B_i^{\rm eff} + \alpha \bm m_i \times (\bm m_i \times \bm B_i^{\rm eff})
\right].
\label{eq:LLGeq}
\end{equation}
Here $\alpha$ is the Gilbert-damping coefficient, which is fixed at $\alpha$=0.1 throughout the present study unless otherwise noted. The effective local magnetic field $\bm B_i^{\rm eff}$ acting on the $i$th magnetization vector $\bm m_i$ is calculated by $\bm B_i^{\rm eff}=-\partial \mathcal{H}/\partial \bm m_i$. The dimensionless time $\tau$ is related with the real time $t$ as $t=\hbar\tau/J$. We investigate the field-induced transformations of magnetic structures among three magnetic states, i.e., elliptical skyrmion, antiskyrmion, and nontopological bubble, confined in a disk-shaped system whose diameter is 71 sites [Figs.~\ref{Fig01}(a)-(c)]. These three magnetic structures have different skyrmion numbers $N_{\rm sk}$, i.e., $N_{\rm sk}$=+1 for the antiskyrmion, $N_{\rm sk}$=$-1$ for the elliptical skyrmion, and $N_{\rm sk}=0$ for the nontopological bubble. 

The skyrmion number $N_{\rm sk}$ is calculated by~\cite{Berg1981,KimJV2020},
\begin{equation}
N_{\rm sk}=\frac{1}{4\pi}\sum_{(i,j,k)}2\tan^{-1}\left[
\frac{\bm m_i \cdot (\bm m_j \times \bm m_k)}
{1 + \bm m_i \cdot \bm m_j + \bm m_j \cdot \bm m_k + \bm m_k\cdot\bm m_i}
\right],
\end{equation}
where $\sum_{(i,j,k)}$ denotes the summation over all the nearest neighbor three-site combinations taken in the counterclockwise order. Note that Eq.~(1) is a standard definition of $N_{\rm sk}$ in the continuum model, while Eq.~(4) is defined on a lattice basis. The definition in Eq.~(1), which involves the spatial derivatives, can lead to unphysical values of $N_{\rm sk}$ when there are sharp spatial variations in the magnetization profile because the accuracy of the finite-difference approximation deteriorates. In contrast, the definition in Eq.~(4) does not involve spatial derivatives and thus provides a more stable and reliable evaluation on a discrete grid.

We adopt $J$(=1) as the energy unit and take $I_{\rm dip}$=0.09, $A$=0.6, and $B_z$=0.015 in the following simulations. The relative stabilities of the three magnetic structures are governed by a competition between the DM interaction and the magnetic dipole interaction. The DM interaction in the crystal with $D_{\rm 2d}$ symmetry stabilizes the antiskyrmion configurations~\cite{Bogdanov1989,Bogdanov1994}. On the contrary, the magnetic dipole interaction stabilizes the skyrmion configurations with closed magnetic-force lines. In Fig.~\ref{Fig01}(d), the calculated energies of the three magnetic structures are plotted as functions of the DM parameter $D$. When $D$ is large, the antiskyrmion is indeed the lowest in energy. On the contrary, when $D$ is small, the skyrmion has the lowest energy because the magnetic dipole interaction acquires relative importance against the DM interaction. The nontopological bubble, which can be regarded as an intermediate magnetic structure of the skyrmion and the antiskyrmion, always has a moderate energy. 

Because of the competition of these two interactions, the skyrmion is elliptically deformed in the $x$ or $y$ direction~\cite{Peng2020,Jena2020,Capic2022}. Specifically, the skyrmion structure is elongated in the $y$ ($x$) direction when it has a clockwise (counterclockwise) helicity when the external magnetic field is perpendicular to the plane. The two skyrmions with opposite helicities are degenerate in energy. The magnetization configuration of the elliptical clockwise skyrmion is shown in Fig.~\ref{Fig01}(a). As seen in this figure, the magnetization vectors on the left and right longitudinal domain walls are aligned in the same orientations as those on the left and right parts of the antiskyrmion in Fig.~\ref{Fig01}(b), which indicates that these magnetization alignments are favored by the DM interaction. The magnetic dipole interaction itself favors a circular skyrmion structure, but the skyrmion spontaneously elongates the longitudinal domain walls to increase the energy gain associated with the DM interaction, which results in the elliptical deformation in the $y$ direction for the clockwise skyrmion. On the contrary, in the case of the counterclockwise skyrmion, the elongation occurs in the transverse direction for the same reason, resulting in the elliptical deformation in the $x$ direction.

The nontopological bubble can be regarded as a combination of a half skyrmion and a half antiskyrmion. The magnetic structure shown in Fig.~\ref{Fig01}(c) has the half skyrmion configuration at its upper portion, while the half antiskyrmion configuration at its lower portion. The half skyrmion part has the magnetization solid-angle sum of $-2\pi$, which corresponds to $N_{\rm sk}=-1/2$, while the half antiskyrmion part has the solid-angle sum of $+2\pi$, which corresponds to $N_{\rm sk}=+1/2$. In total, the nontopological bubble has a zero solid-angle sum and $N_{\rm sk}=0$ due to their cancellation. Importantly, this magnetic structure is directional. We define the orientation of this bubble structure as a direction pointing from the half-antiskyrmion portion to the half-skyrmion portion. Therefore, the direction of the bubble structure in Fig.~\ref{Fig01}(c) is $+\bm y$, and thus we refer to this magnetic structure as $+\bm y$-oriented nontopological bubble. Note that the in-plane components of magnetization vectors constituting the half-skyrmion part are oriented in the same manner as those constituting the half-antiskyrmion part. Hence, this magnetic structure is stabilized by the in-plane magnetic field because both parts can acquire additive Zeeman-energy gains from the in-plane magnetic field $\bm B_{\rm in}$. Specifically, the $+\bm y$-oriented nontopological bubble in Fig.~\ref{Fig01}(c) is stabilized by $\bm B_{\rm in}$$\parallel$$+\bm x$.

A recent experiment for a noncentrosymmetric dipolar magnet Mn$_{1.4}$Pd$_{0.9}$Pt$_{0.1}$Sn demonstrated that the antiskyrmions and the elliptical skyrmions are stable or metastable in its thin-plate sample~\cite{Peng2020,Jena2020}. Furthermore, it was revealed that the nontopological bubbles cannot be the lowest-energy state as far as the external magnetic field is perpendicular to the sample plane, and they cannot be stable unless an in-plane magnetic field is applied. Considering these facts as well as the energy diagram in Fig.~\ref{Fig01}(d), we adopt $D=0.03$ in the present work, for which all the three magnetic structures are close in energy. Starting from one of the three magnetic structures for $D=0.03$ as an initial state, we study possible switching and transformation phenomena induced by the application of an in-plane magnetic field.

Here, it should be noted that the system is assumed to behave as a hard magnet, because our model is designed for systems like Mn$_{1.4}$Pt$_{0.9}$Pd$_{0.1}$Sn, which exhibit helical magnetic structures even at zero magnetic field. We numerically calculated the energy associated with the magnetic dipole interaction and that associated with the perpendicular magnetic anisotropy in the saturated state. We obtained a quality factor $Q$ of approximately 1.58. This result supports the assumption that the system is a hard magnet.

Another thing to be mentioned is the impact of dimensionality. We employ a purely two-dimensional model for the present theoretical study. On the contrary, the samples of Mn$_{1.4}$Pt$_{0.9}$Pd$_{0.1}$Sn used for the experiments in Ref.~\cite{Peng2020} are three-dimensional, although they are thinned significantly for Lorentz transmission electron microscopy observation and can be regarded as a quasi two-dimensional system. It is known that in three-dimensional systems, Landau surface domains are likely to be formed due to the magnetic dipole interaction. Moreover, interlayer magnetic dipole interactions may obscure essential behaviors of the topological magnetic textures and the switching phenomena. Thus, we adopt a two-dimensional model as it is best suited to extract the fundamental physics behind the switching phenomena at the initial stage of theoretical analysis, through avoiding possible complexities of the three-dimensional systems such as surface effects and interlayer interactions. However, for future studies, it will be important to extend the model to three dimensions in order to quantitatively validate our results against experimental systems and to assess the effects of spatial dimensionality more precisely.

\section{Results}
\subsection{Transformation from Antiskyrmion to Nontopological Bubble}
\begin{figure}[tb]
\includegraphics[scale=0.5]{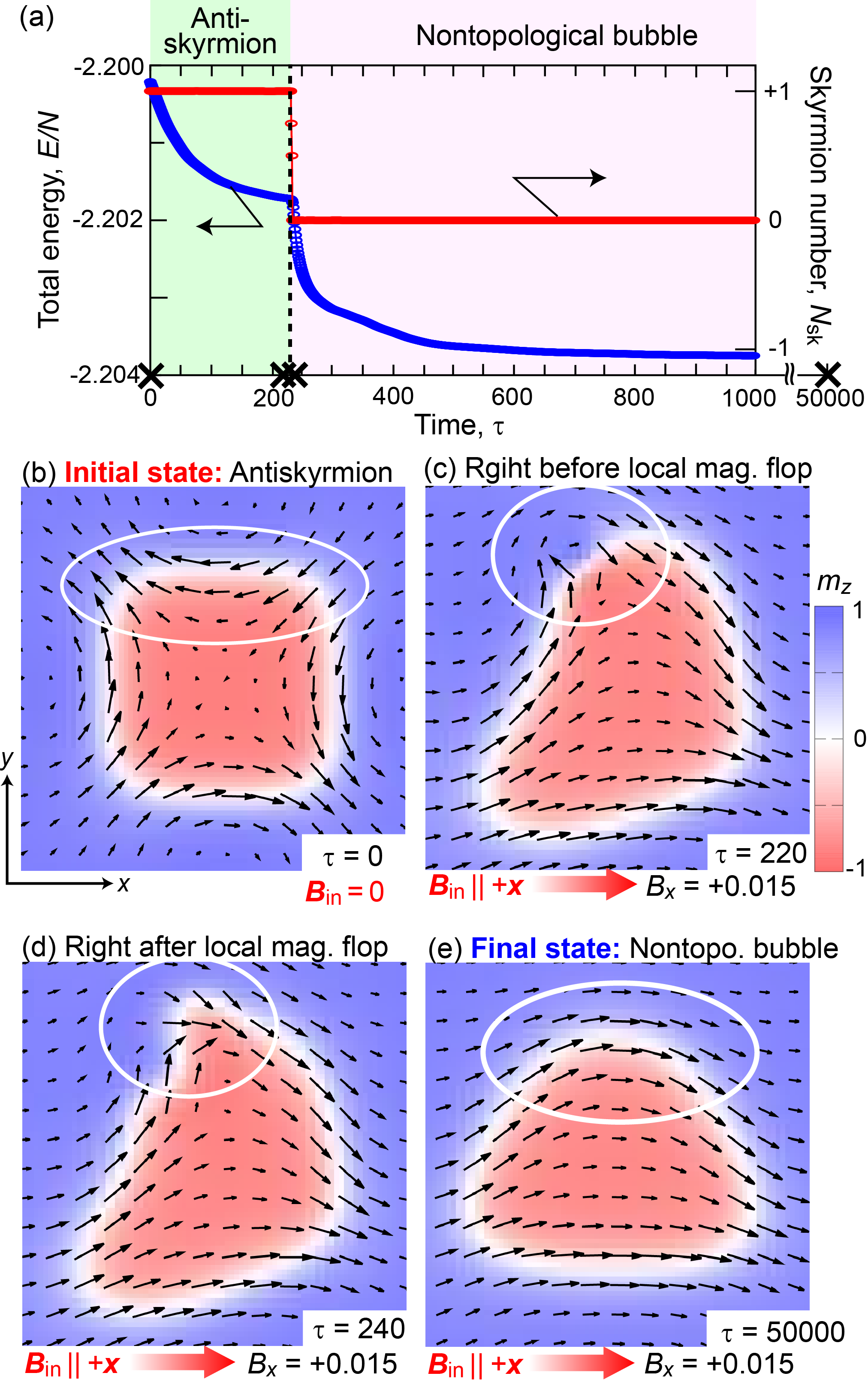}
\caption{(a) Time profiles of the total energy $E$ and the skyrmion number $N_{\rm sk}$ during the transformation from antiskyrmion to nontopological bubble induced by application of an in-plane magnetic field $\bm B_{\rm in}$$\parallel$$+\bm x$. with $B_x=+0.015$. 
(b)-(e) Snapshots of the magnetization configurations at selected moments, i.e., (b) initial antiskyrmion structure, (c) structure right before the local magnetization flop, (d) structure right after the local magnetization flop, and (e) final nontopological-bubble structure after sufficient time has passed. The areas of 36$\times$36 sites are magnified. The corresponding moments are marked with cross symbols on the abscissa in (a).}
\label{Fig02}
\end{figure}
First, we study the case that an in-plane magnetic field along the $+x$ axis, i.e., $\bm B_{\rm in}$$\parallel$$+\bm x$, is applied to the initial-state antiskyrmion [Figs.~\ref{Fig02}(a)-(e)]. The simulated time profiles of the total energy $E$ and the skyrmion number $N_{\rm sk}$ are shown in Fig.~\ref{Fig02}(a). Here the strength of the in-plane magnetic field is fixed at $B_x$=+0.015 in the simulation. After the field application starts at $\tau$=0, the energy decreases monotonically and shows an abrupt change at $\tau$=233. At this moment, $N_{\rm sk}$ exhibits a discontinuous change from +1 to 0, which indicates the switching of magnetic topology due to the transformation from the antiskyrmion to the nontopological bubble. 

Figures~\ref{Fig02}(b)-(e) show the simulated magnetic structures at selected moments during this transformation process. The arrows represent the in-plane components of magnetization vectors, while the colors represent their out-of-plane components. Significant changes in the magnetization configurations appear within areas indicated by solid ellipses. Initially, the magnetization vectors of the antiskyrmion oppose the in-plane magnetic field in the upper area indicated by the solid ellipse in Fig.~\ref{Fig02}(b). They begin to realign themselves to be parallel to the in-plane magnetic field with $B_x (>0)$. This realignment of the magnetization vectors proceeds in a continuous manner for a while because of the topological constraint. After a while, most of the magnetization vectors are oriented along the in-plane magnetic field $\bm B_{\rm in}$$\parallel$$+\bm x$, but a very small area with opposite magnetization survives so as to preserve the topological number [Fig.~\ref{Fig02}(c)]. When the energy in this small area is extremely increased, the magnetization vectors undergo an abrupt reversal so as to resolve this unstable situation with a locally increased energy [Fig.~\ref{Fig02}(d)]. After the reversal, the system is relaxed to the equilibrium nontopological bubble state after sufficient duration [Fig.~\ref{Fig02}(e)]. Note that the magnetic structure in Fig.~\ref{Fig02}(e) may seem to be slightly different from that of the nontopological bubble in Fig.~\ref{Fig01}(c). However, the structure in Fig.~\ref{Fig02}(e) would indeed become the nontopological bubble shown in Fig.~\ref{Fig01}(c) through relaxation after turning off the in-plane magnetic field.

\begin{figure}[tb]
\includegraphics[scale=1.0]{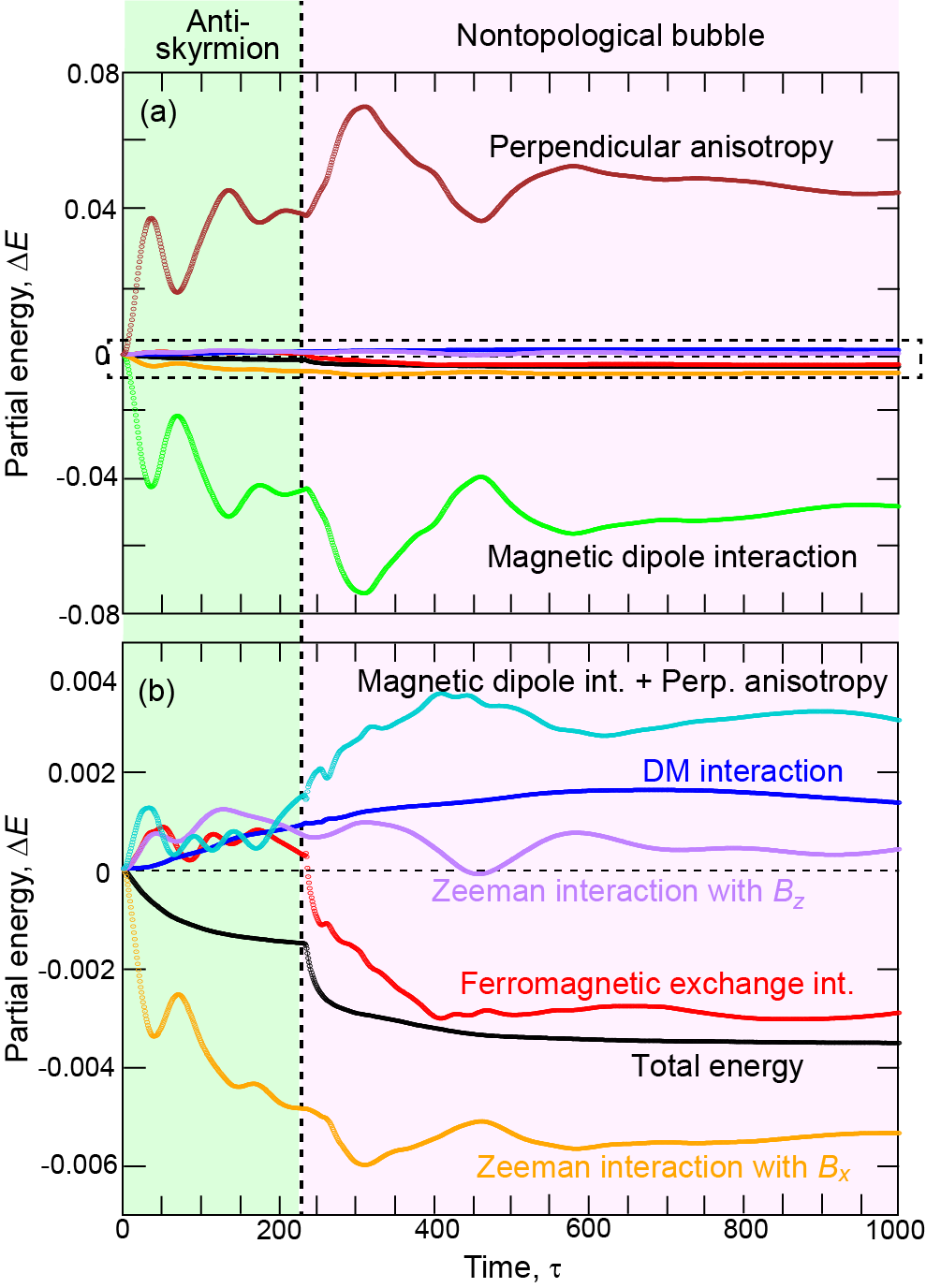}
\caption{(a),~(b) Time profiles of partial contributions to the energy from respective terms of the Hamiltonian for the transformation process from the antiskyrmion to the nontopological bubble under application of the in-plane magnetic field  $\bm B_{\rm in}$$\parallel$$+\bm x$ with $B_x=+0.015$. The values relative to those at $\tau$=0 are plotted. (b) The region around $\Delta E \approx 0$ in (a) is magnified.}
\label{Fig03}
\end{figure}
To clarify the microscopic mechanism of this field-induced switching of magnetic topology, we calculate the time profiles of partial energy contributions from respective terms of the Hamiltonian. Figures~\ref{Fig03}(a) and (b) show the contributions relative to those at $\tau$=0. In Fig.~\ref{Fig03}(a), we find that the partial energies associated with the magnetic dipole interaction and the perpendicular magnetic anisotropy show notable changes. Moreover, it is found that they show bilateral or symmetrical behaviors with similar amplitudes but opposite signs. This is because the magnetic dipole interaction works as an in-plane magnetic anisotropy and cancel the contribution from the perpendicular magnetic anisotropy, which makes it easier for the magnetization vectors to be oriented along the applied in-plane magnetic field. Consequently, the transformation process turns out to be governed by subtle competitions among other interactions.

Figure~\ref{Fig03}(b) magnifies the profiles around $\Delta E$=0 in Fig.~\ref{Fig03}(a), which shows the subtle competitions of other interactions. We find several remarkable behaviors in this figure. First, the energy of the Zeeman interaction associated with the in-plane magnetic field component $B_x$ decreases right after the in-plane field application starts. Second, on the contrary to this behavior, other partial energy contributions from, e.g., the ferromagnetic exchange interaction, the DM interaction, and the Zeeman interaction with the out-of-plane magnetic field component $B_z$ increase in the initial process before the local magnetization flop occurs. Third, the total energy decreases rapidly right after the local magnetization flop occurs. Fourth, this rapid decrease is mostly attributable to the decreasing contribution from the ferromagnetic exchange interaction.

\begin{figure}[tb]
\includegraphics[scale=0.5]{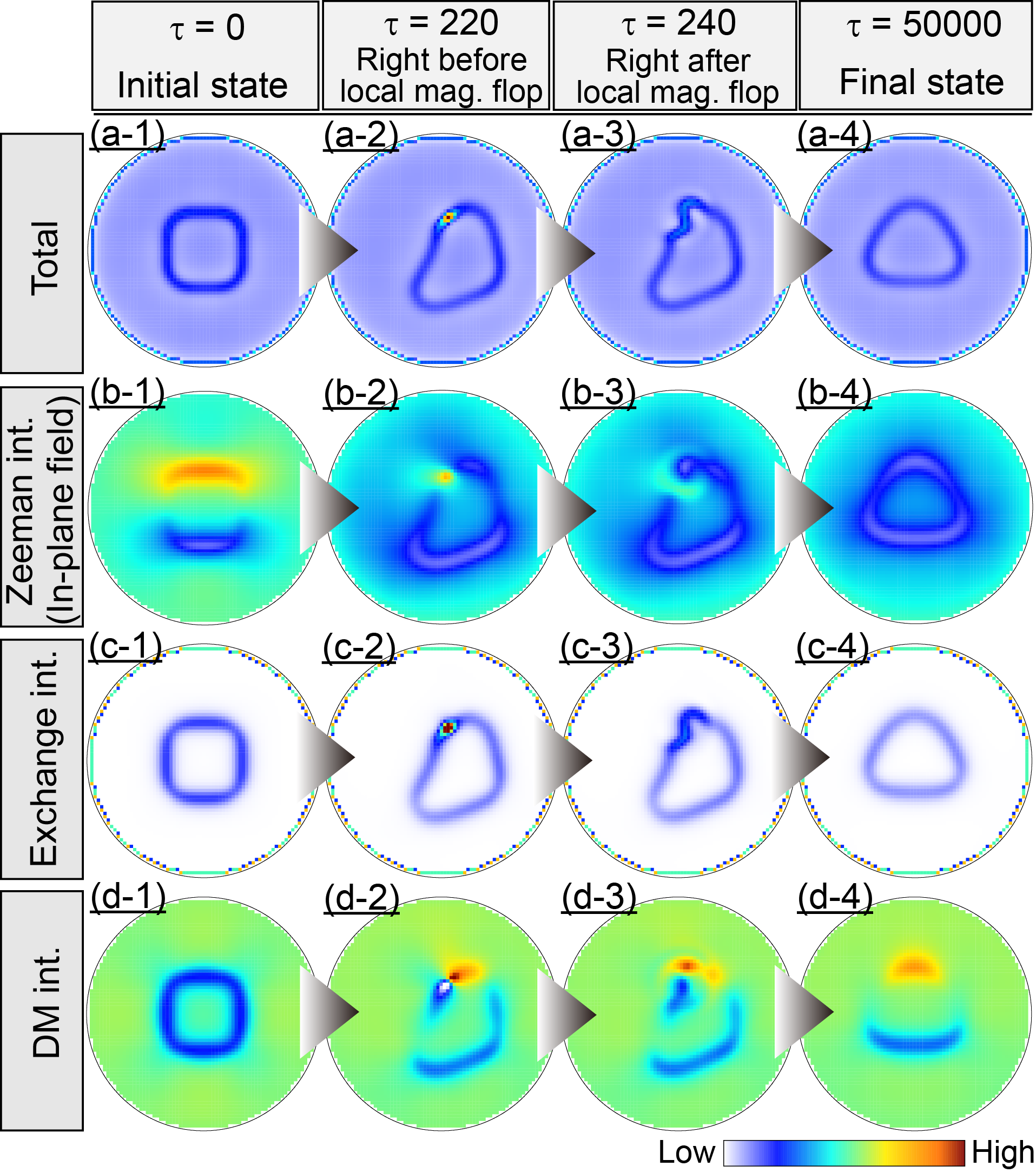}
\caption{(a) Snapshots of the spatial profile of energy density at selected moments during the transformation process from the antiskyrmion to the nontopological bubble under application of the in-plane magnetic field $\bm B_{\rm in}$$\parallel$$+\bm x$ with $B_x=+0.015$. (b)-(d) Those of partial energy densities associated with (b) the Zeeman interaction with the in-plane magnetic field component $B_x$, (c) the ferromagnetic exchange interaction, and (d) the DM interaction.}
\label{Fig04}
\end{figure}
Figure~\ref{Fig04}(a) shows simulated spatial distributions of the total energy density, while Figs.~\ref{Fig04}(b)-(d) show those of the partial energy densities of respective interactions, i.e., (b) the Zeeman interaction with the in-plane magnetic field $B_x$, (c) the ferromagnetic exchange interaction, and (d) the DM interaction. The four panels in each row correspond to selected moments during the transformation process, that is, (1) the initial state ($\tau$=0), (2) a moment right before the local magnetization flop, (3) a moment right after the local magnetization flop, and (4) the final state after long duration. 

In Fig.~\ref{Fig04}(a-2), we find that there appears a small region at which the energy density is extremely high just before the local magnetization flop. This high energy-density region suddenly disappears after the local magnetization flop occurs as seen in Fig.~\ref{Fig04}(a-3). Indeed, this local enhancement of the energy density can be attributed to the Zeeman interaction with the in-plane magnetic field $B_x (>0)$ [Fig.~\ref{Fig04}(b-2)] and the ferromagnetic exchange interaction [Fig.~\ref{Fig04}(c-2)]. In the intial state at $\tau=0$, the upper area of the antiskyrmion is higher in partial energy density of the Zeeman interaction associated with $B_x$ as seen Fig.~\ref{Fig04}(b-1). This is because the magnetizations in this area are oriented in the $-x$ direction and thus are opposite to the in-plane magnetic field of $B_x (>0)$. After a while, the energy density is reduced in most parts of the upper area since gradual realignment of magnetizations occurs. However, because of topological protection, a small area with opposite magnetizations remains, at which the Zeeman interaction associated with $B_x$ and the ferromagnetic exchange interaction have large energy cost. This local energy cost triggers the local magnetization flop and the resulting magnetic topology switching. Interestingly, the energy density of the DM interaction does not show any significant change on the local magnetization flop as shown in Fig.~\ref{Fig04}(d-1)-(d-4), which indicates that the DM interaction does not play major roles in the observed topology switching.

In fact, this physical mechanism is similar to that argued in a previous study in Ref.~\cite{Salikhov2021}, in which the magnetization switching process induced by application of an in-plane magnetic field to stripe domains in [Co/Pt]$_x$ multilayers was observed through measurements of the in-plane magnetization components. In this work, using micromagnetic simulations, it was demonstrated that this switching is triggered by the collapse of magnetic domain walls known as horizontal Bloch lines. This switching mechanism, involving the formation of singularities in the magnetization texture due to the collapse of horizontal Bloch lines, closely resembles the field-induced magnetic topology switching discussed here.

\subsection{Transformation from Skyrmion to Nontopological Bubble}
\begin{figure}[tb]
\includegraphics[scale=0.5]{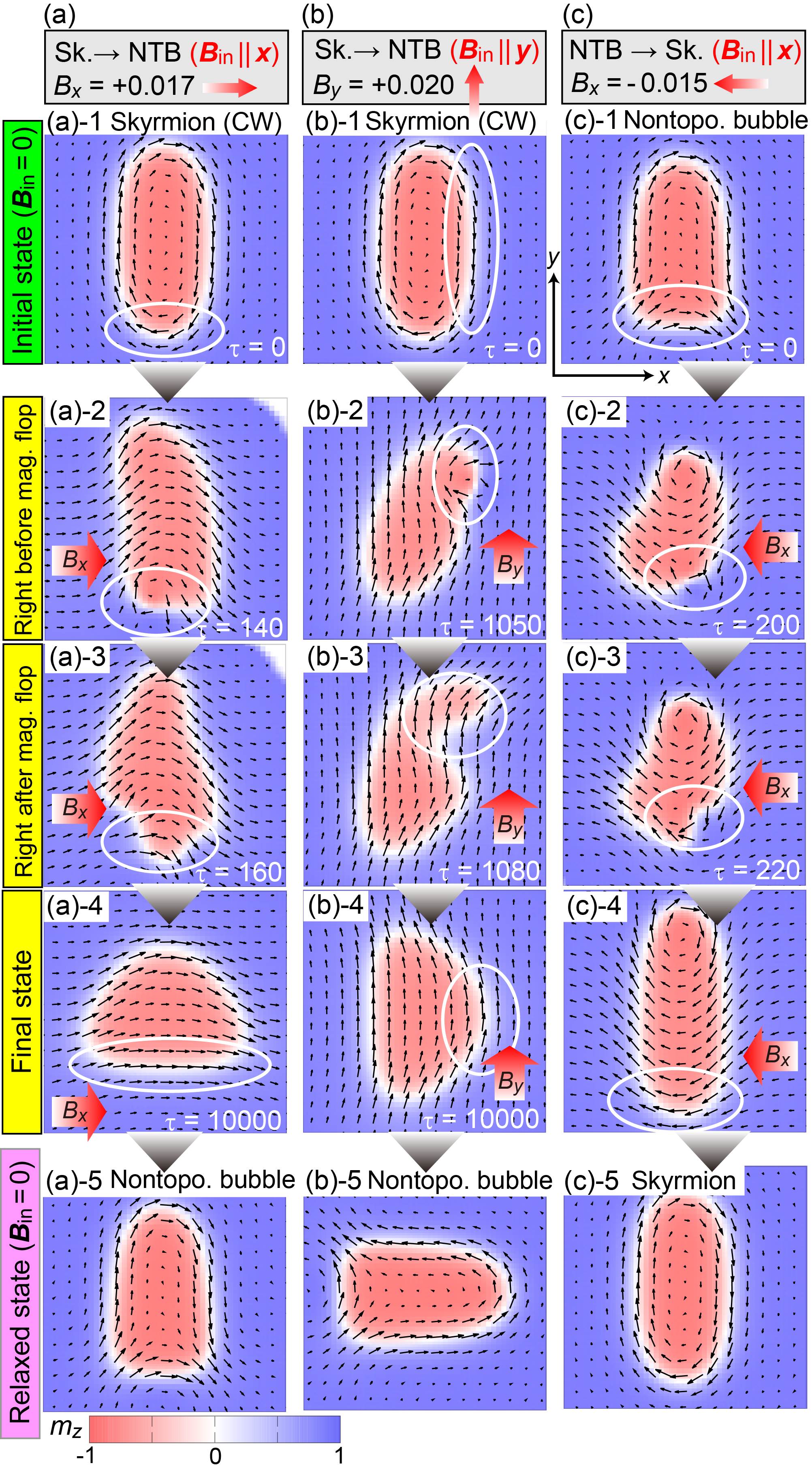}
\caption{Snapshots of the magnetization configurations at selected moments for several types of skyrmion-bubble transformations induced by an in-plane magnetic field, i.e., (a) skyrmion to $+\bm y$-oriented nontopological bubble with $B_x=+0.017$, (b) skyrmion to $+\bm x$-oriented nontopological bubble with $B_y=+0.02$, and (c) nontopological bubble to skyrmion with $B_x=-0.015$. The relaxed states after turning off the in-plane magnetic field $\bm B_{\rm in}$ are also shown in the bottom panels. The areas of 40$\times$40 sites around the magnetic structures are magnified.}
\label{Fig05}
\end{figure}
In addition to the field-induced transformation from the antiskyrmion to the nontopological bubble, we also find several other types of transformations. One example is transformations between the skyrmion and the nontopological bubble (see Fig.~\ref{Fig05}). Starting from the initial-state clockwise skyrmion configuration [Figs.~\ref{Fig05}(a-1) and (b-1)], we obtain the $+\bm y$-oriented and $+\bm x$-oriented nontopological bubble [Figs.~\ref{Fig05}(a-4) and (b-4)] by application of in-plane magnetic fields with $B_x$ and $B_y$, respectively. Importantly, the orientation of the obtained bubble structure depends on the direction of the applied in-plane magnetic field. Namely, a triangular-shaped bubble, base of which is parallel to the $x$ ($y$) axis is obtained when the in-plane magnetic field $\bm B_{\rm in}$$\parallel$$\bm x$ ($\bm B_{\rm in}$$\parallel$$\bm y$) is applied. 

Importantly, this skyrmion-bubble transformation occurs in a reversible manner. As shown in Fig.~\ref{Fig05}(c), we can achieve a transformation from the nontopological bubble to the skyrmion with an in-plane magnetic field. Snapshots of the magnetization configurations at selected moments during the resepective transformation processes [Fig.~\ref{Fig05}] indicate that the physical mechanisms of these magnetic transformations are similar to the mechanism of the antiskyrmion-bubble transformation in Fig.~\ref{Fig02} argued above. Namely, the applied in-plane magnetic field causes a local enhancement of the energy in the course of gradual realignment of the magnetization vectors, which are originally oriented opposite to the applied in-plane field. The resulting energetically unstable situation leads to an abrupt flop of the local magnetization and eventually the switching of magnetic topology.

\subsection{Switching the Orientation of Nontopological Bubble}
\begin{figure*}[tb]
\includegraphics[scale=0.5]{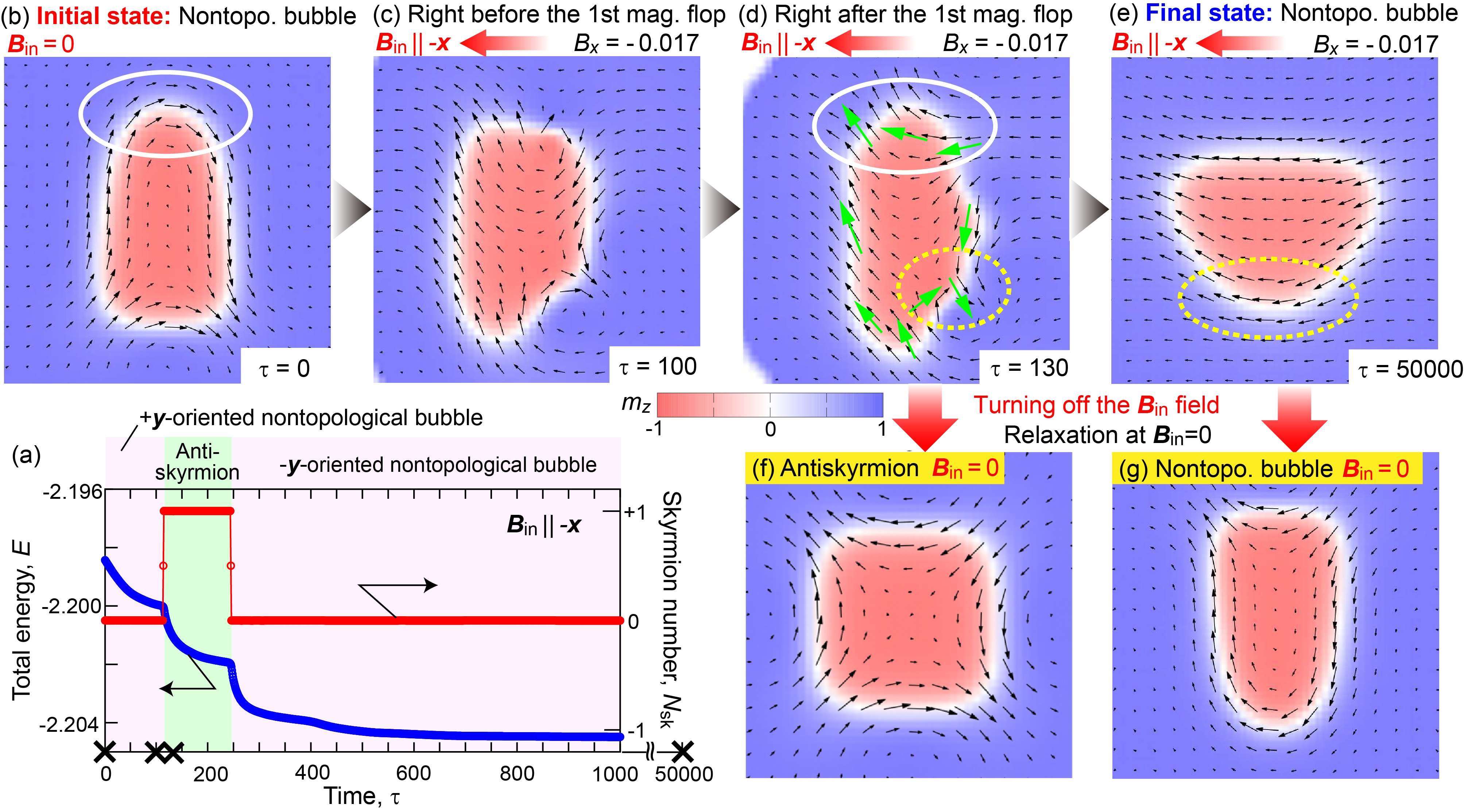}
\caption{(a) Time profiles of the total energy $E$ and the skyrmion number $N_{sk}$ during the successive transformation process from the $+\bm y$-oriented nontopological bubble to the $-\bm y$-oriented nontopological bubble via the antiskyrmion induced by application of an in-plane magnetic field $\bm B_{\rm in}$$\parallel$$-\bm x$ with $B_x=-0.017$. (b)-(e) Snapshots of the magnetization configurations at selected moments, i.e., (b) initial $+\bm y$-oriented nontopological bubble, (c) structure right before the first local magnetization flop, (d) structure right after the first local magnetization flop, and (e) final $-\bm y$-oriented nontopological bubble after sufficient time has passed. The areas of $40\times40$ sites are magnified. The corresponding moments are marked with cross symbols on the abscissa in (a). (f) Antiskyrmion configuration obtained through relaxation when the $\bm B_{\rm in}$ field is turned off at the moment of (d). (g) Nontopologial bubble configuration obtained through relaxation when the $\bm B_{\rm in}$ field is turned off at the moment of (e).}
\label{Fig06}
\end{figure*}
Another example is the field-induced successive transformation process from a nontopological bubble to another nontopological bubble with different orientation via the antiskyrmion [Fig.~\ref{Fig06}], which is observed when we apply an in-plane magnetic field to the initial-state nontopological bubble. Starting from this initial bubble state, we simulate the magnetization dynamics under application of an in-plane magnetic field $\bm B_{\rm in}$$\parallel$$+\bm x$ with $B_x=-0.017$. The calculated time profile of the total energy $E$ in Fig.~\ref{Fig06}(a) shows a monotonic decrease with two anomalies. The time profile of the skyrmion number $N_{\rm sk}$ shows discontinuous changes at these anomaly points among quantized numbers from $0$ to $+1$ to $0$, which indicates that successive transformations from a nontopological bubble [Fig.~\ref{Fig06}(b)] to the antiskyrmion [Fig.~\ref{Fig06}(c)] to another nontopological bubble [Fig.~\ref{Fig06}(d)]. 

Note that as demonstrated in Fig.~\ref{Fig05}(c), the initial-state bubble structure can also be transformed to a skyrmion instead of the antiskyrmion by application of an in-plane magnetic field in the same direction. In that case of Fig.~\ref{Fig05}(c), however, the applied in-plane magnetic field is a little smaller in magnitude as $B_x=-0.015$, for which the local magnetization flop and the local energy enhancement occur dominantly in the lower-half antiskyrmion part. Consequently, the local magnetization flop occurs at a single small area in the lower half part only, which results in the observed single bubble-to-skyrmion transformation in Fig.~\ref{Fig05}(c). On the other hand, we apply a little stronger in-plane magnetic field of $B_x=-0.017$ in the same direction for the present case of the successive transformation process. Figure~\ref{Fig06}(d) shows that the magnetization flop occurs first in the upper-half skyrmion area indicated by solid ellipses in Figs.~\ref{Fig06}(b) and (c), which results in the formation of the deformed antiskyrmion structure with $N_{\rm sk}$=+1 in Fig.~\ref{Fig06}(d). 

When we stop applying the in-plane magnetic field $\bm B_{\rm in}$ at this moment, the deformed antiskyrmion structure is relaxed to be a symmetric antiskyrmion structure shown in Fig.~\ref{Fig06}(e). This  means that we can achieve the bubble-to-antiskyrmion transformation by a controlled application of the in-plane magnetic field. On the contrary, when we continue to apply the $\bm B_{\rm in}$ field, the subsequent local magnetization flop occurs in the area indicated by (yellow) dashed ellipse in Fig.~\ref{Fig06}(d), resulting in the second magnetic topology switching from $N_{\rm sk}$=+1 to $N_{\rm sk}$=0. As a result, we obtain another nontopological bubble structure, which is distinct from the initial bubble structure with respect to the orientation and helicity. The bubble structrure in Fig.~\ref{Fig06}(e) obtained under application of the $\bm B_{\rm in}$ field changes its structure to be that in Fig.~\ref{Fig06}(g) through relaxation after turning off the in-plane magnetic field.

\begin{figure}[tb]
\includegraphics[scale=0.5]{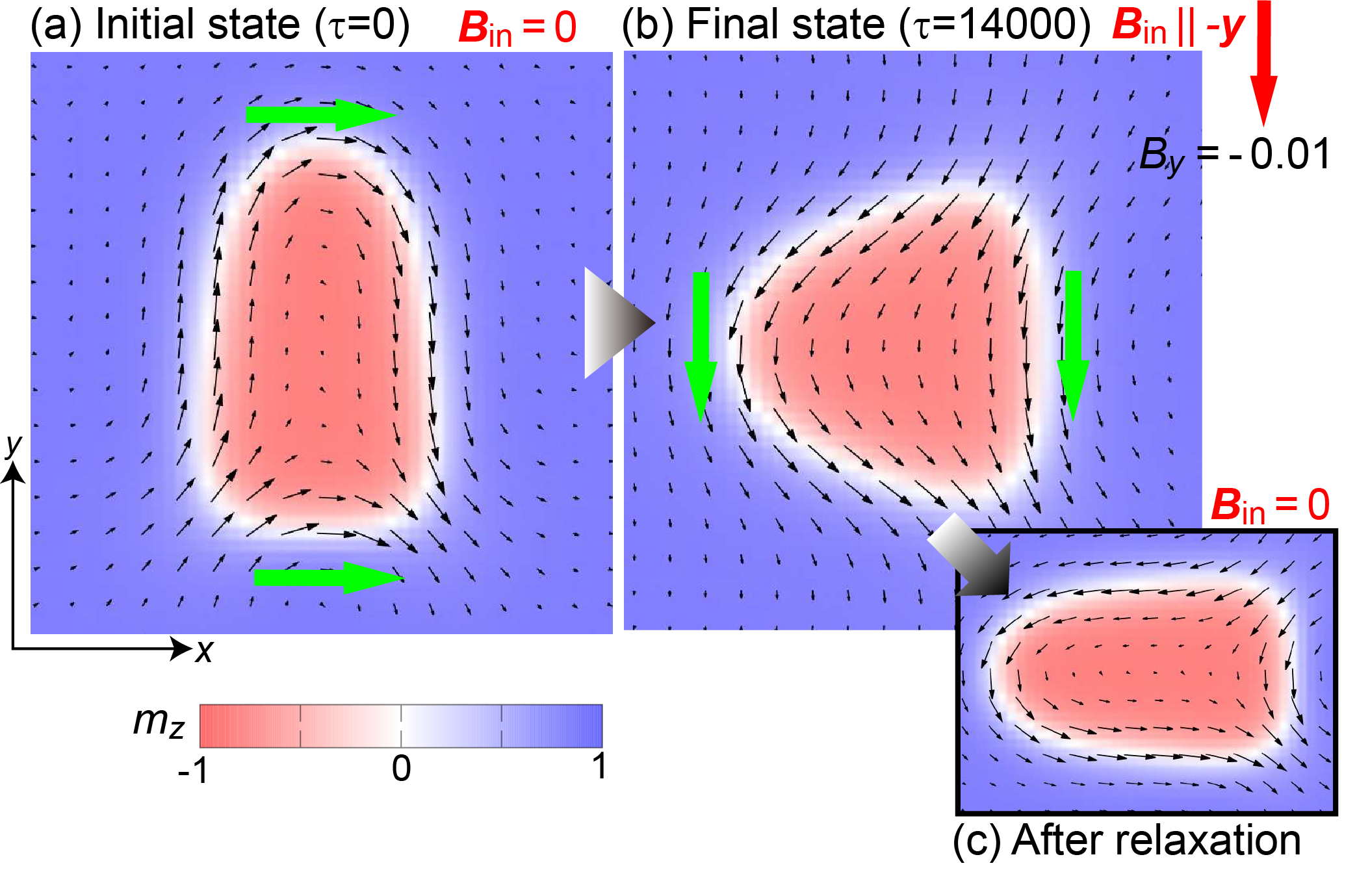}
\caption{90-degree reorientation of the nontopological bubble induced by an in-plane magnetic field $\bm B_{\rm in}$$\parallel$$-\bm y$ with $B_y=-0.01$. (a) Initial $+\bm y$-oriented nontopological bubble at $\bm B_{\rm in}$=0. (b) $-\bm x$-oriented nontopological bubble after the reorientation  occurs. The areas of $40\times40$ sites are magnified. (c) Nontopological bubble structure relaxed sufficiently after the $\bm B_{\rm in}$ field is turned off.}
\label{Fig07}
\end{figure}
The successive transformation process shown in Fig.~\ref{Fig06} can be regarded as a reorientation of the nontopological bubble from the $+\bm y$-oriented to the $-\bm y$-oriented one. In addition to this 180-degree reorientation, we can also realize the 90-degree reorientation as shown in Fig.~\ref{Fig07}. Here an in-plane magnetic field $\bm B_{\rm in}$$\parallel$$-\bm y$ is applied to the initial $+\bm y$-oriented nontopological bubble [Fig.~\ref{Fig07}(a)]. We observe a 90-degree reorientation toward the $-\bm x$-oriented bubble [Fig.~\ref{Fig07}(b)] in the numerical simulations. The bubble structure in Fig.~\ref{Fig07}(b) is relaxed to be that in Fig.~\ref{Fig07}(c) after sufficient relaxation in the absence of the in-plane magnetic field, i.e., $\bm B_{\rm in}$=0. A comparison bwteen Figs.~\ref{Fig07}(a) and (c) indicates that although the orientation of the bubble structure, that is, the direction pointing from the half-antiskyrmion part to the half-skyrmion part is rotated by 90 degrees upon this 90-degree flop, it is found that the spatial profile of magnetization vectors in Fig.~\ref{Fig07}(c) cannot be obtained by the simple 90-degree rotation. The magnetic structure in Fig.~\ref{Fig07}(c) is produced by operating a 90-degree rotation and mirror reflection with respect to the $y$ axis to that in Fig.~\ref{Fig07}(a). We can switch the orientation of nontopological bubble at will by applications of in-plane magnetic fields through combining the 180-degree reversal and the 90-degree flop.

\subsection{Controlled Switching of Magnetic Textures}
\begin{figure*}[tb]
\includegraphics[scale=0.5]{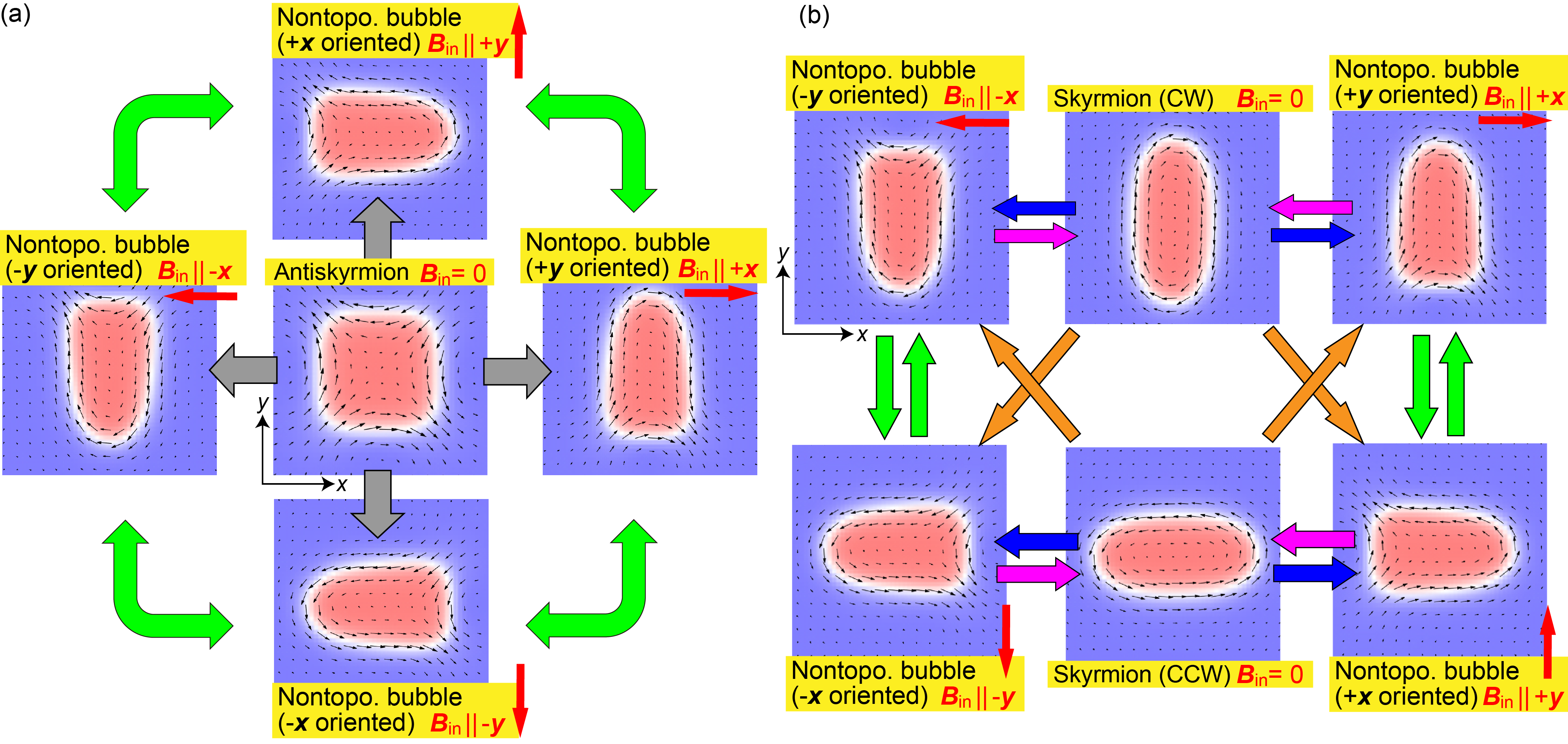}
\caption{Predicted mutual transformations among the antiskyrmion, skyrmions (clockwise and anticlockwise), and nontopological bubbles ($\pm\bm x$- and $\pm\bm y$-oriented) by application of in-plane magnetic fields. (a) Conversions of the antiskyrmion structure to nontopological bubbles with different orientations (gray arrows) as well as switchings of the orientation of nontopological bubble (green arrows). The gray arrows correspond to the process shown in Fig.~\ref{Fig02}, while the green arrows correspond to the process shown in Fig.~\ref{Fig07}. (b) Conversions between the skyrmion structures and the nontopological bubbles (blue, orange and purple arrows) as well as the switchings of the orientation of nontopological bubble (green arrows). The blue and orange arrows correspond to the processes shown in Figs.~\ref{Fig05}(a) and (b), respectively, while the purple arrows correspond to the process shown in Fig.~\ref{Fig05}(c). This diagram shows that the helicity of skyrmion can also be switched between the counterclockwise (CCW) and clockwise (CW) via the nontopological-bubble structure although it cannot be directly changed to each other.}
\label{Fig08}
\end{figure*}
The results of our numerical simulations argued in this paper indicate that we can switch the magnetic structures with different topological numbers $N_{\rm sk}$ in a controlled manner in the noncentrosymmetric dipolar magnets with competing DM interaction and magnetic dipole interaction. The magnetic structures include the antiskyrmion with $N_{\rm sk}$=+1, skyrmions with $N_{\rm sk}$=$-1$, and nontopological bubbles with $N_{\rm sk}$=0. Moreover, there are several types of skyrmions and nontopological bubbles. For skyrmions, there are two types with different helicities, i.e., the clockwise skyrmion elongated in the $y$ direction and the counterclockwise skyrmion elongated in the $x$ direction. These two types of skyrmions are degenerated in energy. On the other hand, there are four types of nontopological bubbles with different orientations, i.e., $\pm \bm x$-oriented and $\pm \bm y$-oriented. These four nontopological bubbles are degenerate in the system with $D_{\rm 2d}$ symmetry, when the external magnetic field is perpendicular to the sample plane without in-plane components. We can realize mutual conversions among these magnetic structures through application of in-plane magnetic fields. 

In Fig.~\ref{Fig08}, we show diagrams of the predicted conversions among antiskyrmion, skyrmions, and nontopological bubbles. The diagram in Fig.~\ref{Fig08}(a) summarizes the conversion from the antiskyrmion structures to nontopological bubbles with different orientations as well as switchings among four nontopological bubbles with different orientations. Here the antiskyrmion-to-nontopological bubble conversions correspond to the process in Fig.~\ref{Fig02}. On the other hand, the orientation switching among the differently oriented four nontopological bubbles corresponds to the process in Fig.~\ref{Fig07}.

The diagram in Fig.~\ref{Fig08}(b) summarizes the predicted conversions among skyrmions and nontopological bubbles in addition to the orientation switching of the nontopological bubble. Here the skyrmion-bubble conversions correspond to processes in Fig.~\ref{Fig05}. Note that the conversions from the (counter)clockwise skyrmion elongated along the $x$ ($y$) axis to the $\pm \bm y$-oriented ($\pm \bm x$-oriented) nontopological bubbles (orange arrows) occur in an irreversible way. Namely, the $\pm \bm y$-oriented ($\pm \bm x$-oriented) nontopological bubbles cannot be converted to the (counter)clockwise skyrmion elongated along the $x$ ($y$) axis purely by application of the in-plane magnetic field. It is worth mentioning that although the helicity of skyrmion cannot be directly switched between the counterclockwise and clockwise skyrmions by the in-plane magnetic field, it can also be switched via the nontopological-bubble structure.

Finally, we compare our results with the experimental results for Mn$_{1.4}$Pd$_{0.9}$Pt$_{0.1}$Sn in Ref.~\cite{Peng2020}. The transformations from antiskyrmion or skyrmion to nontopological bubbles observed in our numerical simulations show excellent agreement with the experimental observations. However, there is one difference between the experimental and theoretical results. Experimentally, it was observed that the nontopological bubbles immediately return to skyrmions when the in-plane magnetic field is turned off. This behavior is not fully reproduced by our simulations. A possible reason for this discrepancy is that the experiment was done for bulk samples at room temperature, whereas our simulations are performed for a nanodisk-shaped sample at zero temperature. Interactions among the magnetic structures via the long-range magnetic dipole interaction and thermal fluctuations are likely responsible for the discrepancy. Therefore, in order to precisely reproduce the experimental results for bulk materials in Ref.~\cite{Peng2020}, it might be necessary to take these factors into account.

\section{Discussion}
In this section, we will discuss several issues to be studied in future research.
\subsection{Effects of thermal fluctuations}
In this study, we have focused primarily on elucidating the mechanism of magnetic topology switching induced by application of an in-plane magnetic field, and therefore the numerical simulations have been performed at zero temperature. This might cause the failure in reproducing the experimentally observed immediate transformation from nontopological bubbles to skyrmions in Mn$_{1.4}$Pd$_{0.9}$Pt$_{0.1}$Sn after turning off the in-plane magnetic field. To reproduce this transformation, effects of thermal fluctuations should be addressed in future work by employing the stochastic Landau-Lifshitz-Gilbert equation with the stochastic magnetic-field term. This approach will enable a quantitative analysis of spontaneous transitions or collapses from the nontopological bubbles to topological states.

However, some caution is required when investigating the effects of thermal fluctuations. The switching of magnetic structures induced by an  in-plane magnetic field can be regarded as a magnetic phase transition driven by the field. In studies of such phase transitions, it is essential to consider dimensionality seriously when dealing with finite-temperature effects. According to the Mermin-Wagner's theorem, which states that no phase transition occurs at finite temperatures in two dimensions with continuous degrees of freedom, indicating that the behavior of phase transitions, the influence of thermal fluctuations, and even the free energy landscape differ fundamentally between two and three dimensions.

Therefore, if one aims to treat finite-temperature effects seriously and explore phenomena related to phase transitions, it is necessary to consider three-dimensional systems, i.e., systems stacked along the thickness direction, instead of purely two-dimensional systems as in this work. However, in systems with the magnetic dipole interaction, simulating such three-dimensional structures is extremely challenging due to high computational costs. We consider the investigation of finite-temperature effects on the magnetic-topology switching phenomena in sufficiently thick three-dimensional systems to be a future task. On the other hand, at zero temperature, the influence of dimensionality is not pronounced. We think that the physical mechanism of the field-induced magnetic-topology switching, which we have elucidated through numerical simulations in two dimensions, is robust and universal.

\begin{figure}[tb]
\includegraphics[scale=0.5]{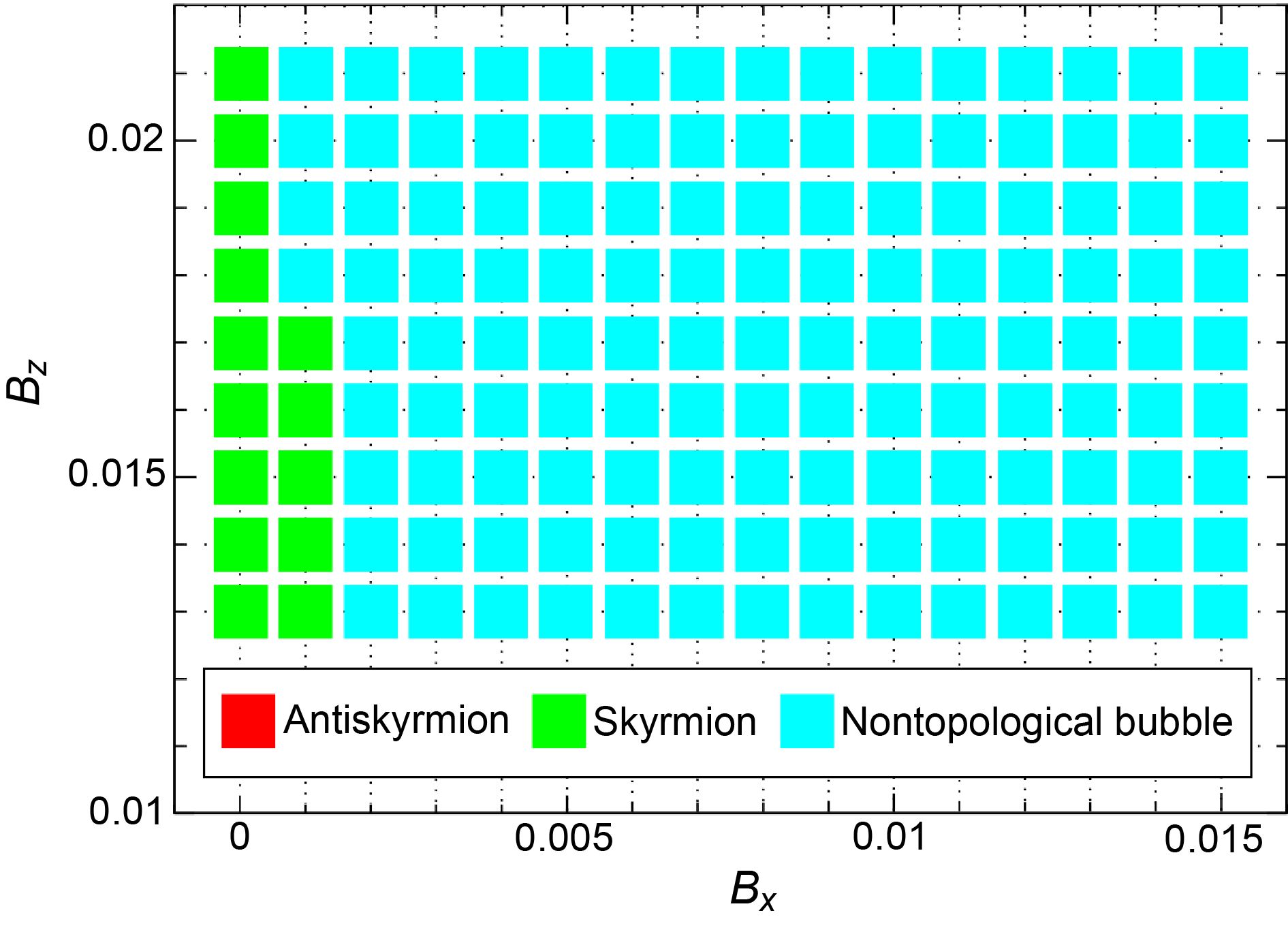}
\caption{Ground-state phase diagram in the plane of the out-of-plane magnetic field $B_z$ and the in-plane magnetic field $B_x$.}
\label{Fig09}
\end{figure}
If we can perform numerical simulations incorporating thermal fluctuations appropriately, it is expected that the transformation from nontopological bubbles to skyrmions observed experimentally upon turning off the in-plane magnetic field can be reproduced over a wide parameter range. In Fig.~\ref{Fig09}, we present the ground-state phase diagram in the plane of the out-of-plane magnetic field $B_z$ and the in-plane magnetic field $B_x$. Here, the other model parameters are set to be the same as for the series of calculations in this study. From this phase diagram, we can see that in the absence of the in-plane magnetic field ($B_x=0$), the skyrmion phase has the lowest energy, whereas even a small in-plane magnetic field ($B_x \ne 0$) stabilizes the nontopological bubble phase. Therefore, by introducing thermal fluctuations at finite temperatures as an assist to overcome the potential barrier, it is anticipated that the transformation between the nontopological bubble state and the skyrmions state can be observed across a wide range of magnetic-field parameters, controlled by turning the in-plane magnetic field on and off. We also note that our simulations reveal that under a strong out-of-plane magnetic field of $B_z > 0.022$, turning off the in-plane magnetic field $B_x$ can induce a transition from the nontopological bubble to the ferromagnetic state instead.

\subsection{Possible bubble-antiskyrmion transformation}
In the experiment reported in Ref.~\cite{Peng2020}, the transition from nontopological bubbles to skyrmions upon turning off the in-plane magnetic field was observed at 250 K. At least at this temperature, the nontopological bubbles change into the skyrmions, whereas no transition to antiskyrmions occurs. This experimental result can be naturally understood from the temperature-field phase diagram shown in Fig.4b of Ref.~\cite{Peng2020}, where the skyrmion phase is the most stable at low temperatures in the absence of the in-plane magnetic field. On the other hand, this phase diagram also indicates that the antiskyrmions become stable in the temperature range above 270 K. Therefore, when the in-plane magnetic field is turned off at temperatures above 270 K, a transition from nontopological bubbles to antiskyrmions is expected to occur. Investigating such temperature-dependent transitions from nontopological bubbles to either skyrmions or antiskyrmions represents an intriguing direction for future research.

\subsection{Evaluation of threshold switching fields}
In this study, we have microscopically elucidated the switching mechanisms. On the other hand, elucidation of the switching conditions, in particular, evaluation of threshold strengths of the in-plane magnetic field for the switching is another important issue. In fact, we quantitatively evaluated the required field strength by conducting numerical simulations through systematically varying the strength of the DM interaction $D$ and the out-of-plane magnetic field $B_z$. We found that the threshold-field strength tends to increase with increasing $B_z$. This tendency can be understood intuitively. The switching of magnetic textures is triggered by singular enhancement of local energy arising when the in-plane magnetization vectors form head-to-head or tail-to-tail configurations. A stronger in-plane magnetic field is neccesary to orient the magnetization vectors within the plane under a stronger $B_z$ field that favors the perpendicular magnetization.

However, we also have found that despite this trend, the evaluated threshold-field strengths are rather scattered and less systematic. This nonsystematic behavior is more pronounced for the $D$-dependence of the threshold field. The lack of systematicity might be attributable to impulsive effects caused by the abrupt application of field in the simulations and/or nonsystematic variations of the demagnetization energy due to the long-range magnetic dipole interactions in the small disk geometry. A quantitative evaluation of the threshold switching field seems to require setups that reflect the real experimental situations, e.g., a gradual application of the in-plane magnetic field to eliminate the impulsive effects and a usage of larger-sized systems to appropriately treat the variation of demagnetization energy. Because such simulations would require high computational costs, we consider a quantitative analysis of the threshold switching field under experimentally realistic conditions to be an issue for future work.

\section{Summary}
In summary, we have theoretically studied the field-induced interconversion phenomena among various topological magnetic textures including antiskyrmion ($N_{\rm sk}$=+1), elliptically deformed skyrmions ($N_{\rm sk}$=$-1$), and nontopological bubbles ($N_{\rm sk}$=0) confined in a nanodisk-shaped sample of noncentrosymmetric dipolar magnet like a Heusler magnet Mn$_{1.4}$Pd$_{0.9}$Pt$_{0.1}$Sn, in which the DM interaction and the magnetic dipole-dipole interaction coexist and are in keen competition. By numerically simulating the spin dynamics induced by applied magnetic fields using the LLG equation, we have demonstrated that a subsequent application of an in-plane magnetic field in addition to the originally applied perpendicular magnetic field can switch the magnetic topology among the above-mentioned magnetic textures in a controlled and deterministic manner. By calculating the profiles of energy contributions from several interactions and magnetic anisotropy, we have revealed the physical mechanism and properties of the observed field-induced topology switching phenomena. We expect that the physics of the field-induced magnetic-topology switching revealed in this study are valid not only for specific inverse-Heusler compounds or $D_{2d}$-symmetric crystals but also for a broader class of materials including schreibersite compounds or those with $S_4$ symmetry~\cite{Karube2021,Karube2022}. Our findings indicate that the noncentrosymmetric dipolar magnets, which have recently been discovered and turned out to host rich topological magnetic textures, provide a promising platform for highly controllable and multifunctional spintronics devices.

\section{Acknowledgments}
This work was supported by JSPS KAKENHI (No.~24H02231 and No.~25H00611), JST CREST (No.~JPMJCR20T1), and Waseda University Grant for Special Research Projects (No.~2024C-153 and No.~2025C-133).

\end{document}